\documentclass[10pt]{iopart}

\usepackage{graphicx}
\usepackage{cite}
\usepackage[dvipsnames]{xcolor}
\usepackage{placeins}

\begin{document}

\title[Shaping the edge radial electric field to create shearless transport barriers in tokamaks]{Shaping the edge radial electric field to create shearless transport barriers in tokamaks}

\author{L A Osorio-Quiroga$^{1,2*}$, M Roberto$^3$, I L Caldas$^1$, \mbox{R L Viana$^4$} and Y~Elskens$^2$}

\address{$^1$University of S{\~a}o~Paulo, Institute of Physics, 05508-090 S{\~a}o Paulo, SP, Brazil}
\address{$^2$Aix-Marseille~Universit{\'e}, UMR 7345 CNRS, PIIM, 13397 Marseille, France}
\address{$^3$Aeronautics Institute of Technology, Physics Department, 12228-900 S{\~a}o Jos{\'e} dos Campos, SP, Brazil}
\address{$^4$Federal University of Paran{\'a}, Physics Department, 81531-990 Curitiba, PR, Brazil}
\ead{$^*$laosorioq1@gmail.com}
\vspace{10pt}
\begin{indented}
\item[]June 2023
\end{indented}

\begin{abstract}
In tokamak-confined plasmas, particle transport can be reduced by modifying the radial electric field. In this paper, we investigate the influence of both a well-like and a hill-like shaped radial electric field profile on the creation of shearless transport barriers (STBs) at the plasma edge, which are a type of barrier that can prevent chaotic transport and are related to the presence of extreme values in the rotation number profile. For that, we apply an $\mathbf{E}\times\mathbf{B}$ drift model to describe test particle orbits \textcolor{black}{in large aspect-ratio tokamaks}. We show how these barriers depend on the electrostatic fluctuation amplitudes and on the width and depth (height) of the radial electric field well-like (hill-like) profile. We find that, as the depth (height) increases, the STB at the plasma edge becomes more resistant to fluctuations, enabling access to an improved confinement regime that prevents chaotic transport. We also present parameter spaces with the radial electric field parameters, indicating the STB existence for several electric field configurations at the plasma edge, for which we obtain a fractal structure at the \textcolor{black}{barrier/non-barrier} frontier, typical of quasi-integrable Hamiltonian systems.
\end{abstract}

\maketitle
\ioptwocol

\section{Introduction \label{sec:Introduction}}

 \textcolor{black}{In tokamaks, modifying the $\mathbf{E}\times\mathbf{B}$ shear can lead to changes in turbulence and transport, enabling access to improved confinement regimes \cite{Biglari1990, burrell1999tests, wagner1, HIDALGO2006679}. In particular, the radial electric field $E_r$ at the plasma edge can be adjusted to reduce particle transport \cite{taylor1989h, weynants1992confinement, wagner1, Devynck2003}. Both the $E_r$-shear and $E_r$-curvature play crucial roles in suppressing turbulence and creating an edge transport barrier \cite{moyer1995beyond, kamiya2016experimental, kobayashi2017turbulent}. For instance, by applying biased-electrode-induced electric fields, modifications of the $E_r$ profile can be made and high-mode-like regimes can be triggered, thereby improving plasma confinement \cite{taylor1989h, van2003turbulent, grenfell2018}. Additionally, these modifications of the radial electric field may occur spontaneously when the neutral beam heating exceeds a certain threshold, leading to the development of a deep well-like structure of $E_r$ inside the last closed flux surface, which results in a transition from a low-confinement to a high-confinement mode \cite{wagner, burrell2004, viezzer2013}. Specifically, as the depth of the radial electric field well increases, the plasma confinement regime tends to improve \cite{mcdermott2009, viezzer2013}.}\par 

In this paper, we show that the presence of such radial electric fields in the plasma edge can cause the formation of shearless transport barriers (STBs) due to the presence of shearless tori. These STBs are basically invariant tori for which the rotation number has local extrema \cite{del1996area}.  Some works have shown that, when the plasma has non-monotonic sheared profiles for the electric and/or the magnetic fields, STBs can appear  \cite{horton1998,caldas2012,del2012gyroaverage,marcus2019,grime2022shearless,morrisonpop}.\par

Our results are obtained from a model of guiding centre motion with ${\bf E}\times{\bf B}$ drift in tokamaks, in which we adopt non-monotonic profiles for the equilibrium radial electric field, and sheared profiles for both the safety factor and the parallel velocity \cite{horton1998}. Although the spectrum of turbulent electrostatic fluctuations is complex, we simplify our model by focusing on a single spatial mode with a finite number of harmonics. Despite this simplification, through our analysis, we observe a correlation between the presence of either a well-like or a hill-like radial electric field profile near the plasma edge and the onset of an STB in tokamaks.\par

\textcolor{black}{Our numerical results indicate that\textcolor{black}{,} by shaping $E_r$ at the plasma edge\textcolor{black}{,} an STB can appear and, thereby, prevent the particles from escaping. Furthermore, we show that the robustness of this STB depends on the profile parameters; in particular, for a \mbox{well-like (hill-like)} $E_r$ profile, the STB will be more resistant to perturbations as the well (hill) \mbox{depth (height)} and width increase.}\par

Since the applied model has a Hamiltonian structure\textcolor{black}{, the phase space flow generated by solving the equations of motion is area-preserving in an adequate Poincar\'e surface of section.} If the plasma profiles were all monotonic, KAM theory would apply everywhere in phase space and shearless barriers would not be possible at all. However, since the profiles are non-monotonic, local extrema correspond to shearless invariant tori which act as dikes, preventing particle diffusion. \par

An undeniable advantage of simple models over large-scale computer simulations is the possibility of choosing a small number of system parameters to investigate the effects of their changes on the STB properties. The depth/height of the radial electric field well/hill-like profile, its width and the intensity of a single mode of the electrostatic potential perturbation can be varied in order to get parameter planes indicating a transport property, namely, the mean escape time for particles. The frontier between escape and non-escape is fractal, which is ultimately the consequence of the complicated invariant curve structure of quasi-integrable Hamiltonian systems. \par

The remainder of this paper is organized as follows: \sref{sec:Drift_Model} outlines the ${\bf E}\times{\bf B}$ drift wave transport model, whereas our numerical results on the identification of shearless transport barriers in the plasma edge are presented in \sref{sec:Edge_Barrier}. In \sref{sec:L_H}, we discuss the influence of the edge radial electric field on the STB robustness, regarded from the point of view of our parameter planes. Finally, we draw our conclusions in \sref{sec:Conclusions}. 

\section{$\mathbf{E}\times\mathbf{B}$ Drift Wave Transport Model\label{sec:Drift_Model}}

For a magnetically confined plasma in a tokamak, let us consider an individual test particle whose guiding centre is moving along the magnetic field lines, $\mathbf{B}(\mathbf{x})$, with velocity $\mathbf{v}_\parallel(\mathbf{x})$ and drifted by a $\mathbf{v}_{\mathbf{E}\times\mathbf{B}}(t,\mathbf{x})$  velocity, such that
\begin{equation} \label{eq:Motion_equation}
	\qquad
	\frac{\rmd\mathbf{x}}{\rmd t} = v_\parallel\hat{b} + \mathbf{v}_{\mathbf{E}\times\mathbf{B}},\quad \mathbf{v}_{\mathbf{E}\times\mathbf{B}} = \frac{\mathbf{E}\times\mathbf{B}}{B^2},
\end{equation}

\noindent where $\mathbf{x}=(r,\theta,\varphi)$ corresponds to the particle position in toroidal coordinates, $v_\parallel=v_\parallel(\mathbf{x})$ its velocity component in the magnetic field direction, \mbox{$\hat{b}=\mathbf{B}/B$}, and $\mathbf{E}=\mathbf{E}(t,\mathbf{x})$ the electric field  experienced by the particle.\par   

\textcolor{black}{For simplicity, we assume} that the magnetic equilibrium surfaces cross sections are concentric circles and that there is only a radial dependence of the plasma profiles. Also, we are ignoring the drifts due to the magnetic field lines curvature and magnetic gradients. To do that, some assumptions are made on the tokamak geometry and the magnetic field; mainly, we take the cylindrical approximation for the tokamak, for which $a/R=\textcolor{black}{\epsilon}\ll 1$, with $a$ and $R$ the minor and major radius of the plasma, respectively, i.e.\ the plasma is treated as a $2\pi R$ periodic cylinder. Furthermore, we assume a screw pinch configuration such that \mbox{$\mathbf{B}(r)=B_\theta\hat{e}_\theta+B_\varphi\hat{e}_\varphi$}, with $B\approx B_\varphi\gg B_\theta$ and $B=\mathrm{cst}$. The radial dependence of the magnetic field components will be regarded through the safety factor profile \textcolor{black}{$q(r)$, given by}

\begin{equation}\label{eq:safety_factor}
\qquad
\textcolor{black}{
q(r) = rB_\varphi(r)/(RB_\theta(r)).
}
\end{equation}
          
The electric field, $\mathbf{E}(t,\mathbf{x})$, is considered as a rotation-free vector field, $\nabla\times\mathbf{E}=0$. When this condition is fulfilled, the electric field can be called electrostatic, even with it depending explicitly on time. In the equilibrium,  we are neglecting any contribution of the parallel electric field, $\mathbf{E}_\parallel$, and \textcolor{black}{considering only} the radial equilibrium part $E_r(r)\hat{e}_r$. For the non-equilibrium scenario, a perturbation is included via the electrostatic potential $\tilde{\phi}(t,\mathbf{x})$, and therefore

\begin{equation}\label{eq:Electric_field}
	\qquad
	\mathbf{E}(t,\mathbf{x}) = E_r(r)\hat{e}_r - \nabla \tilde{\phi}(t,\mathbf{x}).
\end{equation}

The electrostatic potential $\tilde{\phi}(t,\mathbf{x})$ is written as a superposition of harmonic waves travelling in the poloidal and toroidal directions,
 
\begin{equation}\label{eq:Electrostatic_potential}
	\qquad
	\tilde{\phi}(t,\theta,\varphi)=\sum_{n}\phi_n\cos(M\theta-L\varphi-n\omega_0t-\alpha_n),
\end{equation}  
	
\noindent where $M$ and $L$ are their dominant wave numbers, respectively, $\omega_0$ their fundamental angular frequency, $\phi_n$ the amplitude and $\alpha_n$ the phase for each perturbation mode.\par

\textcolor{black}{Now,} on using two new variables, \textcolor{black}{the action and the angle, defined, respectively, as}

\begin{equation}
    \textcolor{black}{\qquad
    \eqalign{
    I = \left(\frac{r}{a}\right)^2, \cr 
    \psi = M\theta - L\varphi,
    }
    }
\end{equation}

\noindent the equations of motion \eref{eq:Motion_equation} reduce to the time dependent one-degree-of-freedom dynamical system

\begin{equation}\label{eq:Dynamical_system}
	\qquad
	\eqalign{
		\frac{\rmd I}{\rmd t}=2M\sum_{n}\phi_n\sin(\psi-n\omega_0t-\alpha_n),\cr
		\frac{\rmd\psi}{\rmd t} = \epsilon v_{\parallel}(I)\frac{[M-Lq(I)]}{q(I)}-\frac{M}{\sqrt{I}}E_{r}(I),}
\end{equation} 

\begin{equation}\label{eq:frequency_profile}
    \qquad
    \omega(I) = \frac{\rmd\psi}{\rmd t},
\end{equation}

 \noindent where we adimensionalize \eref{eq:Motion_equation} using the characteristic scales $a$, \mbox{$E_a=|E_r(a)|$} and $B$ according to the relations 

\begin{equation}\label{eq:Adimensionalize}
	\qquad
	\eqalign{	
	E_r'=\frac{E_r}{E_a}, \quad \phi_n'=\frac{\phi_n}{aE_a}, \quad v_\parallel'=\frac{B}{E_a}v_\parallel, \cr
	t'=\frac{E_a}{aB}t, \quad \omega_0'=\frac{aB}{E_a}\omega_0,} 
\end{equation}

\noindent \textcolor{black}{and $\omega(I)$ is the angular frequency of the motion.} Note that \textcolor{black}{in equations \eref{eq:Dynamical_system} and \eref{eq:frequency_profile}} the prime notation was omitted.\par

\textcolor{black}{According to \eref{eq:Dynamical_system}, the radial particle transport will be mainly governed by the electrostatic potential perturbation, $\tilde{\phi}(t,\psi)$, and the particle rotation by the plasma radial profiles: $E_r(I)$, $v_\parallel(I)$, and $q(I)$. In particular, these profiles can alter the poloidal rotation of the plasma, for instance, through \mbox{$\mathbf{E}_r\times\mathbf{B}_\varphi$}, corresponding to the last term of the angular equation of motion.}

\textcolor{black}{The variables $I$ and $\psi$ represent the action-angle canonical set of the unperturbed Hamiltonian}  

\begin{equation}\label{eq:Hamiltonian_0}
    \qquad
    \textcolor{black}{
    H_0(I) = \int^I \omega(I')\,\rmd I',
    }
    \end{equation}

\noindent \textcolor{black}{whereas the system \eref{eq:Dynamical_system} describes the evolution of these variables when a perturbation $H_1(t,\psi)$ is introduced, resulting in the perturbed Hamiltonian $H(t,\psi, I)$, given by}

\begin{equation}\label{eq:Hamiltonian}
    \qquad
    \textcolor{black}{
    H(t,\psi, I) = H_0(I) + H_1(t,\psi),
    }
    \end{equation}

\noindent \textcolor{black}{where}
    
    \begin{equation}\label{eq:Hamiltonian_1}
    \qquad
    \textcolor{black}{
    \eqalign{
    H_1(t,\psi) = 2M\tilde{\phi}(t,\psi), \cr
    \frac{\rmd I}{\rmd t} = -\frac{\partial H}{\partial \psi}, \quad \frac{\rmd \psi}{\rmd t} = \frac{\partial H}{\partial I}.
    }
    }
\end{equation}

\textcolor{black}{We notice that if $M = 0$, i.e. the perturbation does not propagate along the poloidal direction, or if $\phi_n=0$ for all modes, the dynamical system will be integrable and fully described by $H_0(I)$. For these scenarios, $I$ remains constant and, consequently, the guiding centre of the test particle traces a helix of constant radius. Now, when we consider the perturbation $H_1(t,\psi)$, the integrability of the system is broken, leading to chaotic behaviour and particle transport outside the plasma, as we show in the next sections.}\par 

The presented model, given by \eref{eq:Dynamical_system}, was introduced in \cite{horton1998}. It considers robust shearless transport barriers \textcolor{black}{(STBs)} which can appear when non-monotonic radial plasma profiles are regarded \cite{horton1998, marcus2019, osorio2021, grime2022shearless}, e.g.\ for the equilibrium electric field\textcolor{black}{, $E_r(I)$}, the parallel velocity\textcolor{black}{, $v_\parallel(I)$}, or the safety factor\textcolor{black}{, $q(I)$}.\par

In this work, we investigate the effect of the radial equilibrium electric field profile on chaotic transport \textcolor{black}{at the plasma edge, $I_{\mathrm{edge}} = 1.0$. Specifically, we} consider that, \textcolor{black}{near $I_{\mathrm{edge}}$}, $E_r(I)$ \textcolor{black}{have either} a \mbox{well-like} \textcolor{black}{or a hill-like} structure. \textcolor{black}{These kinds of profiles can be found, for instance,} \textcolor{black}{in tokamaks operating in an} \mbox{H-mode regime \textcolor{black}{\cite{asdex1989,grenfell2018}}}. As we know, in the H-mode, close to the edge, a pedestal structure appears in the plasma density, pressure, and temperature profiles \cite{wagner,connor2000}, which reflects in the typical radial electric field profile \cite{viezzer2013,ida1990edge,grenfell2018} and in the appearance of a transport barrier \cite{wagner1984,horton2005characterization, grenfell2018}.\par 

\FloatBarrier

\section{Shearless Edge Transport Barriers\label{sec:Edge_Barrier}}

We construct a numerical map by integrating the dynamical system \eref{eq:Dynamical_system} and considering the solution at times $T_j=2j\pi/\omega_0$, with \mbox{$j=0,1,2,3,\dots, N$}. This procedure defines a $\psi\times I$ stroboscopic Poincar\'e phase portrait, that, given an initial condition \mbox{$\mathbf{P}_0=(\psi_0,I_0)$}, will describe the regular or \textcolor{black}{chaotic} orbit \mbox{$\Sigma_N=(\mathbf{P}_0,\mathbf{P}_1,\dots,\mathbf{P}_N)$}.  The results were obtained using the numerical integrator \mbox{Runge-Kutta-Dormand-Prince} of 8(7) order \cite{prince1981high}, \textcolor{black}{which is a numerical integrator method of order $\mathcal{O}(h^8)$ that uses an error estimate of order $\mathcal{O}(h^7)$ to control the adaptive step size, $h$\cite{engeln1996numerical}. The Poincar\'e phase portraits presented in this section were obtained using an error tolerance of $10^{-13}$.}\par 

\begin{figure}[htb]
	\begin{center}
		\includegraphics[width = 0.48\textwidth]{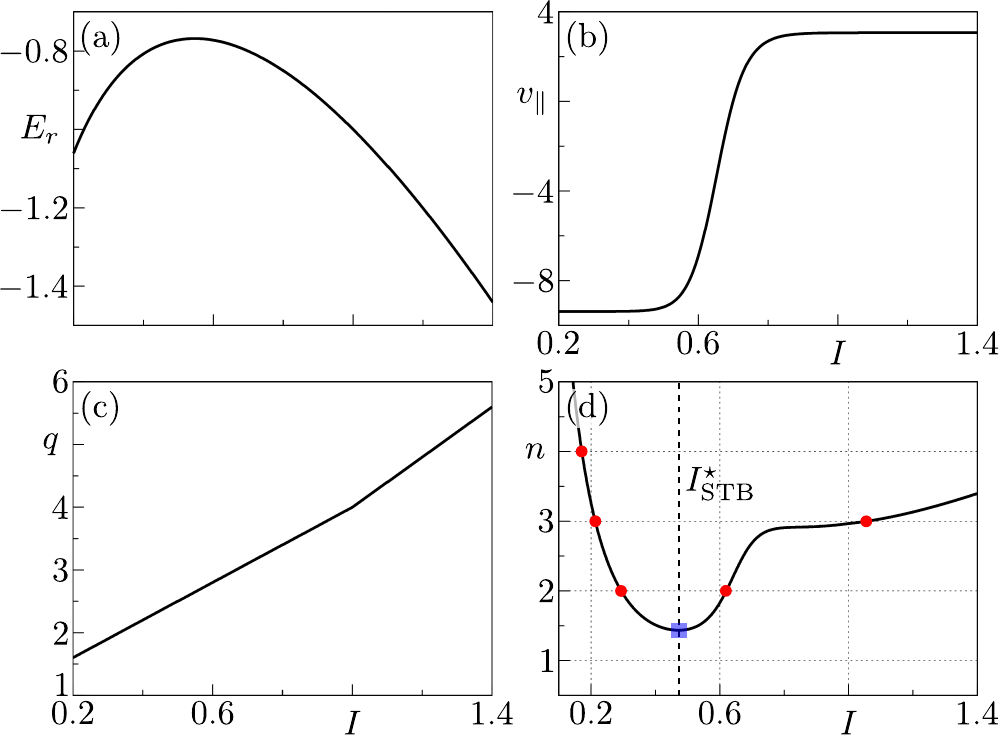}
	\end{center}
	\caption{\label{fig:profiles}Plasma radial profiles for the (a) equilibrium radial electric field, $E_r(I)$, (b) parallel velocity, $v_\parallel(I)$, and (c) safety factor, $q(I)$. In panel (d), we show the resonance conditions profile, indicating in red \textcolor{black}{dots} the resonant modes we are considering for the \textcolor{black}{electrostatic potential perturbation}, \mbox{$n=2,3~\mathrm{and}~4$}. The amplitude $\phi_1$, which is related in this figure to the non-resonant mode $n=1$, is taken as one of the control parameters. \textcolor{black}{The vertical dashed line and the blue square represent the shearless point position, $I_{\mathrm{STB}}^{\star}$, of the profile.}}
\end{figure}

Additionally, we take into account the plasma profiles and parameters for the tokamak TCABR, mainly, the radial equilibrium electric field \cite{marcus2019}, $E_r(I)$, the parallel velocity \cite{severo2003plasma, rosalem2016}, $v_\parallel(I)$, and the safety factor \cite{fernandes2016instabilidades}, $q(I)$, which are specified in the equations \eref{eq:Plasma_profiles} \textcolor{black}{and \eref{eq:Electric_field_profile}}, and in \fref{fig:profiles}, respectively. For TCABR, the minor and major plasma radii are \mbox{$a=0.18$ m} and \mbox{$R=0.61$ m}, respectively, the tokamak minor radius \mbox{$b=0.21$ m} and the toroidal magnetic field \mbox{$B=1.20$ T}. Moreover, the characteristic  scale of the electric field is taken as \mbox{$E_a=2.274$ kV/m}. Explicitly,
 
\begin{equation}\label{eq:Plasma_profiles}
	\qquad
	\eqalign{
		E_r(I) = \textcolor{black}{E_{\mathrm{L}}(I)}, \cr
        v_\parallel(I) = \chi + \zeta\tanh(\xi I^{1/2} + \kappa),\cr
		q(I)= \left\{ \begin{array}{ccc}
			\rho+\varsigma I & \mathrm{if} & I \leq 1 \cr
			(\rho+\varsigma)I & \mathrm{if} & I > 1,
		\end{array}
		\right.}
\end{equation}

\noindent where

\begin{equation}\label{eq:Electric_field_profile}
\qquad
    \textcolor{black}{E_{\mathrm{L}}(I) = 3\alpha I+2\beta I^{1/2} + \gamma}
\end{equation}

\noindent and all the greek letters are dimensionless parameters which are kept fixed for the purpose of this work. They correspond to $\alpha=-1.14$, $\beta=2.53$, \mbox{$\gamma=-2.64$}, \mbox{$\chi=-3.16$}, $\zeta=6.22$, $\xi=20.30$, $\kappa=-16.42$, $\rho=1.00$ and $\varsigma=3.00$. \par

In relation to the electrostatic potential \textcolor{black}{perturbation} parameters, based on the experimental data analysis made in \cite{grenfell2016estudo}, we take for the fundamental angular frequency  $\omega_0=60$ \textcolor{black}{rad$/$ms} (approximately \mbox{$5.70$ rad} after carrying out the adimensionalization), and as dominant spatial modes, $M=16$ and $L=3$. The phase constant is kept for all modes as $\alpha_n=\pi$.\par

The resonance conditions for the dynamical system \eref{eq:Dynamical_system} are given by 

\begin{equation}\label{eq:Resonant_condition}
	\qquad
	\eqalign{
		\frac{\rmd}{\rmd t}\left(\psi-n\omega_0t-\alpha_n \right) = 0,\cr
		\mathrm{i.e}. \quad n = \frac{\omega(I)}{\omega_0}.}
\end{equation}

When considering all the plasma profiles and parameters already defined above, we obtain the resonance conditions\textcolor{black}{, or the angular \textcolor{black}{frequency ratio}, see \eref{eq:Resonant_condition}, radial profile} indicated in \fref{fig:profiles}(d). Here, we mark with red dots the main resonant modes $n=2,3$ and $4$; any contribution \textcolor{black}{of $\phi_n$} for $n>4$ is neglected. From the figure, we note the modes $n=2$ and $n=3$ are resonant in two positions, while \mbox{$n=4$} just in one. The electrostatic potential \textcolor{black}{perturbation amplitudes} for each of these modes are taken, respectively, as \mbox{$0.80$ V}, \mbox{$1.50$ V} and \mbox{$0.85$ V}, which become the dimensionless fixed parameters \mbox{$\phi_2=1.95\times10^{-3}$}, $\phi_3=3.66\times10^{-3}$ and \mbox{$\phi_4=2.08\times10^{-3}$}.\par

\begin{figure*}
	\begin{center}
		\includegraphics[width = 0.9\textwidth]{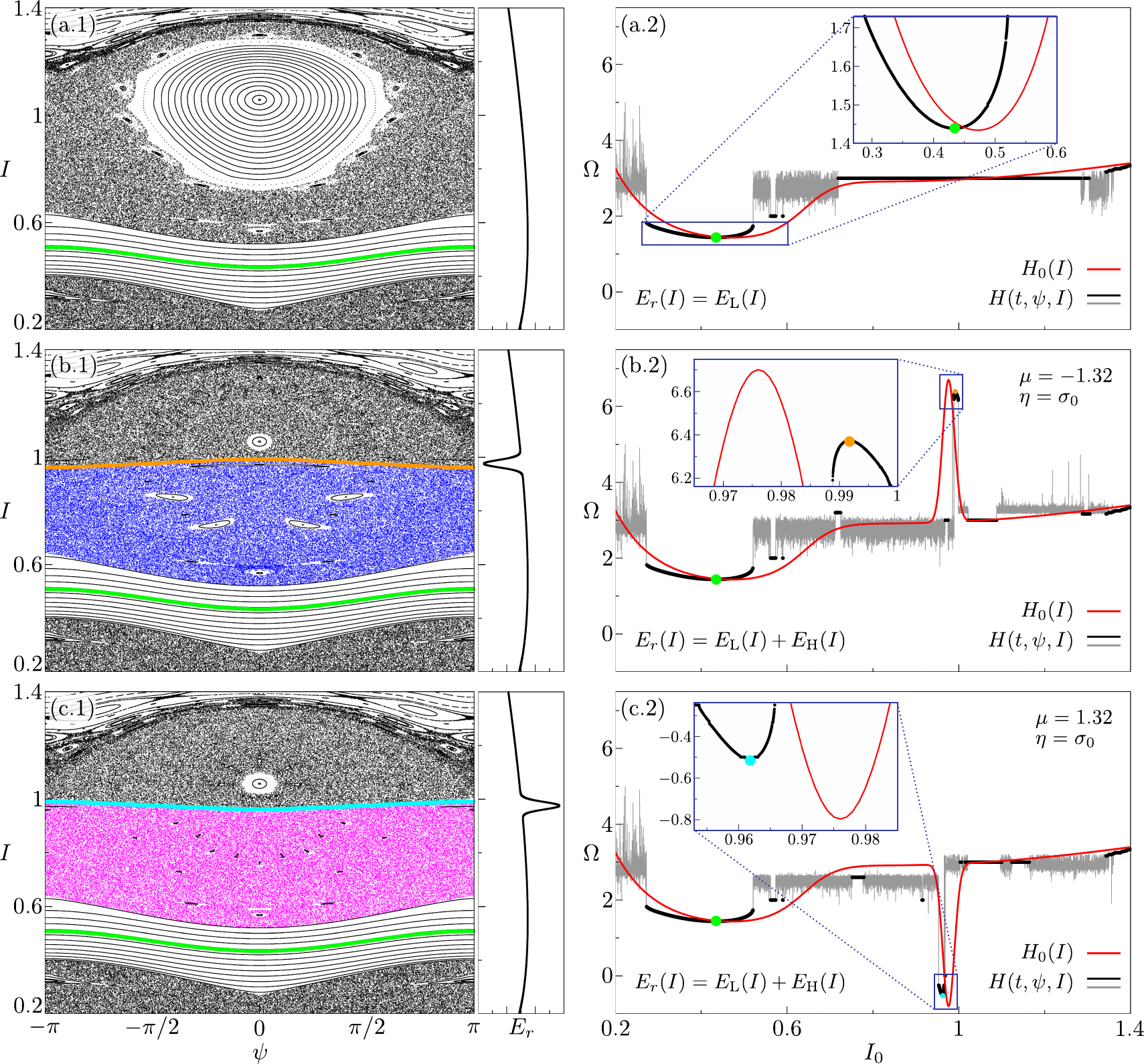}
	\end{center}
	\caption{\label{fig:phase_space} Poincar\'e sections (left panels) and rotation number profiles (right panels) for (a) $E_r(I) = \textcolor{black}{E_{\mathrm{L}}(I)}$, \mbox{(b) $\textcolor{black}{E_r(I) = E_{\mathrm{L}}(I) + E_{\mathrm{H}}(I)}$} \textcolor{black}{with $\eta=\sigma_0=7\times10^{-3}$ and \mbox{$\mu=-1.32$}}, and (c) \mbox{$\textcolor{black}{E_r(I) = E_{\mathrm{L}}(I) + E_{\mathrm{H}}(I)}$} with \textcolor{black}{$\eta=\sigma_0$ and \mbox{$\mu=1.32$}. For the three scenarios $\phi_1 = 0.0$, \mbox{$\phi_2=1.95\times10^{-3}$}, $\phi_3=3.66\times10^{-3}$ and \mbox{$\phi_4=2.08\times10^{-3}$}.} To the right of each Poincar\'e section we show the radial equilibrium electric field profile which was used in each case. Shearless transport barriers are highlighted in \textcolor{black}{green (inner), orange (edge) and \mbox{cyan (edge)} on the Poincar\'e sections. They \textcolor{black}{are} identified through the shearless points, marked with filled circles according to their corresponding colour, in the rotation number profiles, for which we fixed $\psi_0 = 0$. Here, we show both cases the unperturbed, in red, and perturbed, in black and grey for regular and non-convergent chaotic orbits, respectively}.}
\end{figure*}

\textcolor{black}{Additionally,} we regard $\phi_1$, which is related to the non-resonant mode $n = 1$ \textcolor{black}{in the former case, see \fref{fig:profiles}(d)}, as one of the control parameters to study the chaotic transport in the plasma edge. \textcolor{black}{As we will demonstrate in Section \ref{sec:L_H}, the amplitude of this perturbation mode serves as a reliable indicator of the STB robustness. Specifically, we will consider the shearless barrier to be strong if it exists for high values of $\phi_1$.} \par

\textcolor{black}{Previous studies \cite{marcus2019,osorio2021} have demonstrated that non-resonant perturbation modes can lead to the emergence or break-up of STBs in a recurrent manner,  even when their amplitudes increase. However, we observe that the influence of $\phi_1$ on transport does not significantly differ when this mode becomes resonant, \textcolor{black}{and} the onset and break-up of the STB remain frequent as $\phi_1$ varies.} \par
     
The amplitude $\phi_1$ is varied taking values in the interval $[0~\mathrm{V}, 15.00~\mathrm{V}]$, i.e. \mbox{$\phi_1 = [0, 3.66\times 10^{-2}]$} after carrying out the adimensionalization. So, the electrostatic perturbation, $\tilde{\phi}$, will oscillate with a maximum amplitude of $18.15$ V, which is consistent with experimental \textcolor{black}{observations} in TCABR\cite{grenfell2016estudo}.\par 
    
The existence of STBs can be associated with the extreme values of the rotation number profile, $\Omega(\psi_0,I_0)$, which \textcolor{black}{can be} determined \textcolor{black}{numerically} by \eref{eq:Rotation_number}, for a fixed initial value  $\psi_0$:

\begin{equation}\label{eq:Rotation_number}
	\qquad
	\eqalign{
		\Omega(\psi_0,I_0) & =  \lim_{N\rightarrow \infty}\frac{1}{2\pi}\frac{1}{N}\sum_{i = 0}^{N-1}(\psi_{i+1}-\psi_{i})\cr
		& =  \lim_{N\rightarrow\infty}\frac{1}{2\pi}\frac{\psi_N(I_0)-\psi_0}{N}.}
\end{equation}

In non-twist systems, STBs can appear and KAM theory does not apply \cite{del1993chaotic, del1996area}. These barriers exhibit more resistance to perturbations than regular KAM tori, whose resistance is related to their rotation number, according to the KAM theorem \cite{reichl1992transition}. Even when an STB is broken up, a stickiness region can emerge preventing the chaotic flux \cite{szezech2012effective, szezech2009transport}.\par  

So, when $(\rmd \Omega(\psi_0,I_0)/\rmd I_0)_{\mathbf{P}_{\mathrm{STB}}}=0$, the KAM assumption is violated and a shearless curve can be identified through the initial condition \mbox{$\mathbf{P}_{\mathrm{STB}}=(\psi_0,I_{\mathrm{STB}})$}. This is why we refer to these types of barriers as ``shearless" \cite{del1996area}. \textcolor{black}{It is important to remark} the rotation number converges only for regular orbits, and it is independent of the choice of $\mathbf{P}_0$, as long as $\mathbf{P}_0$ lies on the orbit. Nevertheless, the radial profile of the rotation number depends on the chosen $\psi_0$ due to the arrangement of orbits on the Poincar\'e section, resulting in different profiles for different $\psi_0$ values. Explicitly, to find the shearless curve, any value of $\psi_0$ can be used \textcolor{black}{since the barrier is a continuous line, so that} there exists at least one point on the barrier for every $\psi$, as \textcolor{black}{the Poincar\'e sections in} \fref{fig:phase_space} illustrate.\par 

\textcolor{black}{The radial profile of the rotation number for $H_0(I)$ can be obtained analytically, it corresponds to the angular \textcolor{black}{frequency ratio} profile $\omega(I)/\omega_0$. From \fref{fig:profiles}(d), we observe that our system is non-twist and has a shearless curve where \mbox{$(\rmd \omega(I)/\rmd I)_{I_{\mathrm{STB}}^{\star}} = 0$}, i.e.\ approximately at $I_{\mathrm{STB}}^{\star} = 0.472$, where the superscript ``$\star$'' is used to distinguish the unperturbed \mbox{scenarios \textcolor{black}{($H_1=0$)}} from the perturbed ones \textcolor{black}{(\mbox{$H_1\neq0$})}. On the other hand, for $H(t, \psi, I)$, the rotation number profile is obtained numerically using \eref{eq:Rotation_number}. It is expected to be similar to $\omega(I)/\omega_0$, except for the non-convergent chaotic regions that emerge after some tori are destroyed. Consequently, if the shearless curve exists for certain value of the perturbation, it is expected to be \textcolor{black}{near} $I_{\mathrm{STB}}^{\star}$, as we show below.}\par

\textcolor{black}{Thus}, let us consider $\phi_1=0$\textcolor{black}{, the amplitudes for modes $n=2,3$ and $4$ as they were set earlier,} and the electric field \textcolor{black}{given} in equation \eref{eq:Electric_field_profile}, see panels (a) of \fref{fig:phase_space}. In this case, we found \textcolor{black}{an} STB, coloured in \textcolor{black}{green, see \fref{fig:phase_space}(a.1)}, using the rotation number profile shown \textcolor{black}{in \fref{fig:phase_space}(a.2). Here, the rotation number profile for $H(t,\psi,I)$, in black and grey for the regular and \mbox{non-convergent} chaotic orbits, respectively, and the profile for $H_0(I)$, in red, are presented. We see that, although some tori are destroyed, the ones which survive compose the non-monotonic part of the rotation number profile by which we identified the STB at $I_{\mathrm{STB}} = 0.434$. This value is close to $I_{\mathrm{STB}}^{\star}$ as we see from the magnification presented in the inset of \fref{fig:phase_space}(a.2).}\par

\textcolor{black}{However,} even though there is a shearless curve, a reasonable fraction of chaotic orbits above it are escaping and reaching the vessel wall, \textcolor{black}{at} \mbox{$I_{\mathrm{wall}}=(b/a)^2=1.36$}\textcolor{black}{, see \fref{fig:phase_space}(a.1)}. Beyond this value, the trajectories lack physical sense; nevertheless, we plot the phase space a little further. \textcolor{black}{Additionally,} also the orbits trapped by the main island are crossing the plasma edge. \textcolor{black}{And, because of this, the current configuration is not satisfactory for particle confinement.}\par

\textcolor{black}{In this article, we show that better confinement configurations can be obtained} if we add, to our current $E_r$ profile, \textcolor{black}{a local pronounced shear \textcolor{black}{reversal} close to $I_{\mathrm{edge}}$. For this}, let us assume a \textcolor{black}{\mbox{well/hill-like} edge} radial electric field profile, $E_\mathrm{H}(I)$, as the one given by

\begin{equation}\label{eq:Electric_field_H-mode}
	\qquad
	E_\mathrm{H}(I) = \mu\exp\left[-\frac{1}{2}\left(\frac{\sqrt{I}-\delta}{\eta}\right)^2\right],     
\end{equation}

\noindent where $\delta = r_{\mathrm{s}}/a = 0.988$ \textcolor{black}{is a} fixed dimensionless parameter \textcolor{black}{which} represents an estimation of the \textcolor{black}{local} electric field \textcolor{black}{shearless point} position. The well \textcolor{black}{(hill)} depth \textcolor{black}{(height)}, $\mu<0$ \textcolor{black}{$(\mu>0)$, and the associated width, $\eta$}, will be treated as control parameters. \par

\textcolor{black}{With that, the new radial equilibrium electric field profile is given by}
\begin{equation}
\qquad
    \textcolor{black}{E_r(I) = E_{\mathrm{L}}(I) + E_{\mathrm{H}}(I).}
\end{equation}

The use of expression \eref{eq:Electric_field_H-mode} rests on the fact \textcolor{black}{that} it is easier to treat the most common parameters associated with \textcolor{black}{H-mode} regimes, such as \mbox{depth \textcolor{black}{(height)}}, width and well \textcolor{black}{(hill)} position of the edge radial electric field. Moreover, it fits the experimental data \cite{sauter2011,viezzer2013,grenfell2018}.\par

\begin{figure}
	\begin{center}
		\includegraphics[width = 0.48\textwidth]{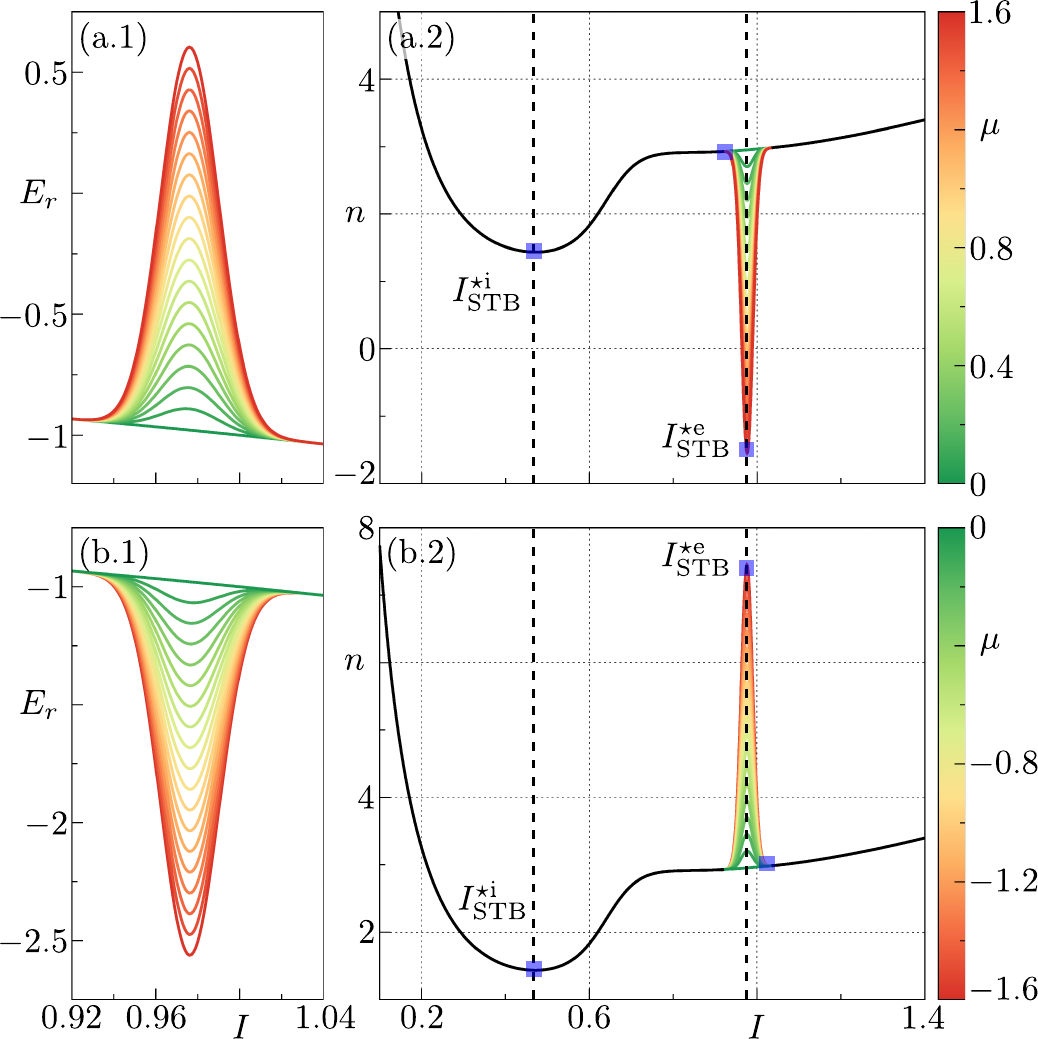}
	\end{center}
	\caption{\label{fig:mu_dependence} \textcolor{black}{As we vary the control parameter $\mu$, (a.1) a \mbox{hill-like} and (b.1) a well-like radial electric field profiles, for $\mu>0$ and $\mu<0$, respectively, and $\eta=\sigma_0=7\times 10^{-3}$, appear near the plasma edge}. At the right panels, (a.2) and (b.2), it is observed  that \textcolor{black}{these profiles} can activate new resonances directly related to the appearance of a shearless \textcolor{black}{edge} transport barrier \textcolor{black}{at $I_{\mathrm{STB}}^{\star\mathrm{e}}$. The point $I_{\mathrm{STB}}^{\star\mathrm{i}}$ corresponds to the internal STB\textcolor{black}{, already indicated in the \fref{fig:profiles}(d) for the former case, $\mu = 0$. The vertical dashed lines and the blue squares represent the shearless point positions of the profile.}}}
\end{figure}

From \fref{fig:mu_dependence}, we observe that the inclusion of this kind of profile will activate new resonant perturbation modes \textcolor{black}{at the plasma edge as $\mu$ is varied}. As a consequence, a new dynamics is induced close to \textcolor{black}{\mbox{$I_{\mathrm{edge}}$}}, which, eventually, will provoke the onset of new shearless curves. \textcolor{black}{\textcolor{black}{Namely, by looking at the right panels of the figure, two} shearless points in each \textcolor{black}{frequency ratio} profile \textcolor{black}{can be noticed}, where in the unperturbed scenarios \textcolor{black}{($H_1 = 0$)} the STBs will appear. \textcolor{black}{For both the well-like and the \mbox{hill-like} profiles,} at $I_{\mathrm{STB}}^{\star\mathrm{i}} = 0.472$  and near the edge at \mbox{$I_{\mathrm{STB}}^{\star\mathrm{e}}=0.976$}, where ``i'' and ``e'' are used to identify the internal and the edge shearless points, respectively. \textcolor{black}{A third shearless point might be found,} to the left of $I_{\mathrm{edge}}$ when $\mu>0$ and to the right of $I_{\mathrm{edge}}$ when $\mu<0$. \textcolor{black}{However, it is not noticeable neither in \fref{fig:mu_dependence} nor in the maps of \fref{fig:phase_space} and} will not be significant in our results.}\par 

\textcolor{black}{Furthermore, notice from the right panels of the figure that the coloured-in-black segments of the \textcolor{black}{frequency ratio} profiles are the same as in the previous case where $\mu=0$, indicating that the internal shearless curve is also the same at $I_{\mathrm{STB}}^{\star\mathrm{i}}$. Then, the inclusion of $\textcolor{black}{E_{\mathrm{H}}}(I)$ will only affect the edge of the plasma, as we will show next. Additionally, notice that for hill-like scenarios the mode $n=1$ can be resonant, explicitly when $\mu \textcolor{black}{\geq} 0.687$.}

In \fref{fig:phase_space}\textcolor{black}{(b.1) and \fref{fig:phase_space}(c.1)},  when \mbox{$\mu=-1.32$} \textcolor{black}{and $\mu=+1.32$, respectively, and \mbox{$\eta=\sigma_0=7\times 10^{-3}$}}, we show that \textcolor{black}{the shearless edge transport barriers, coloured in orange and cyan, respectively, exist and confine most of} the orbits inside the plasma. So, \textcolor{black}{for the \mbox{well-like (hill-like)} $E_r$ profile,} the chaotic blue \textcolor{black}{(magenta)} orbit cannot go through the orange \textcolor{black}{(cyan)} shearless curve\textcolor{black}{, which already \textcolor{black}{indicates} being an improved confinement regime for plasma. The shearless internal transport barriers are coloured in green, as in the previous scenario.}\par

\textcolor{black}{These barriers were found using} the rotation number profiles shown in \fref{fig:phase_space}(b.2) and \fref{fig:phase_space}(c.2). There, we observe that\textcolor{black}{, in comparison with \fref{fig:phase_space}(a.2), where $\mu=0$, the inner part of the profiles (i.e. for $I<0.6$ approximately) is unchanged and does not depend on $\mu$. The invariant tori and internal STBs are the same in the three cases, as we can verify from the Poincar\'e sections. The appearance of} shearless edge transport barriers does not affect the dynamics closer to the plasma centre. \textcolor{black}{Furthermore, as in the previous case, the rotation number profiles of the perturbed scenarios, in black and grey, are similar to the unperturbed ones, in red. At the plasma edge, we found that \mbox{$I_{\mathrm{STB}}^{\mathrm{e}}=0.992$} when $\mu = -1.32$ and that \mbox{$I_{\mathrm{STB}}^{\mathrm{e}}=0.962$} when \mbox{$\mu = +1.32$}.}\par

\textcolor{black}{Moreover, we see that,} in order to allow the edge transport barrier to appear, the main resonance\textcolor{black}{, see \fref{fig:phase_space}(a.1)}, is shrunk, \textcolor{black}{as shown in} \fref{fig:phase_space}(b.1) \textcolor{black}{and \fref{fig:phase_space}(c.1). Here, the outer tori are destroyed and the ones which survive, closer to the centre of the island, conserve their} rotation number \textcolor{black}{$\Omega = 3.0$. By looking} at the rotation number profiles, \textcolor{black}{it is \textcolor{black}{noticeable} that, for $\mu = 0$, there is a large plateau which is shrunk approximately in the interval \mbox{$1.02<I_0<1.09$} when $|\mu|=1.32$. Here, instead of a large plateau, there are non-convergent chaotic regions and the non-monotonic part of the rotation number profile by which we identify the shearless edge transport barriers in both cases}.\par

\textcolor{black}{Now,} a natural question remains after observing the existence of the indicated barrier: does the \textcolor{black}{edge STB} become more resistant \textcolor{black}{to perturbations} as  $|\mu|$ increases? Experimental observations have shown that one of the main distinctions between the \mbox{L-mode}, \mbox{I-mode} and \mbox{H-mode} is the radial electric field intensity at the plasma edge since the electric well structure is deeper as the confinement regime improves \cite{viezzer2013}. To answer this question, we surveyed the parameter space $\mu\times\phi_1$ to identify when the barrier can appear or be broken up. \textcolor{black}{We study, as well, the influence of the width, $\eta$, of the $\textcolor{black}{E_{\mathrm{H}}}(I)$ profile on the shearless edge transport barrier robustness.}
\FloatBarrier

\section{Influence of the edge radial electric field on chaotic transport \label{sec:L_H}}

\begin{figure}[htb]
	\begin{center}
		\includegraphics[width = 0.48\textwidth]{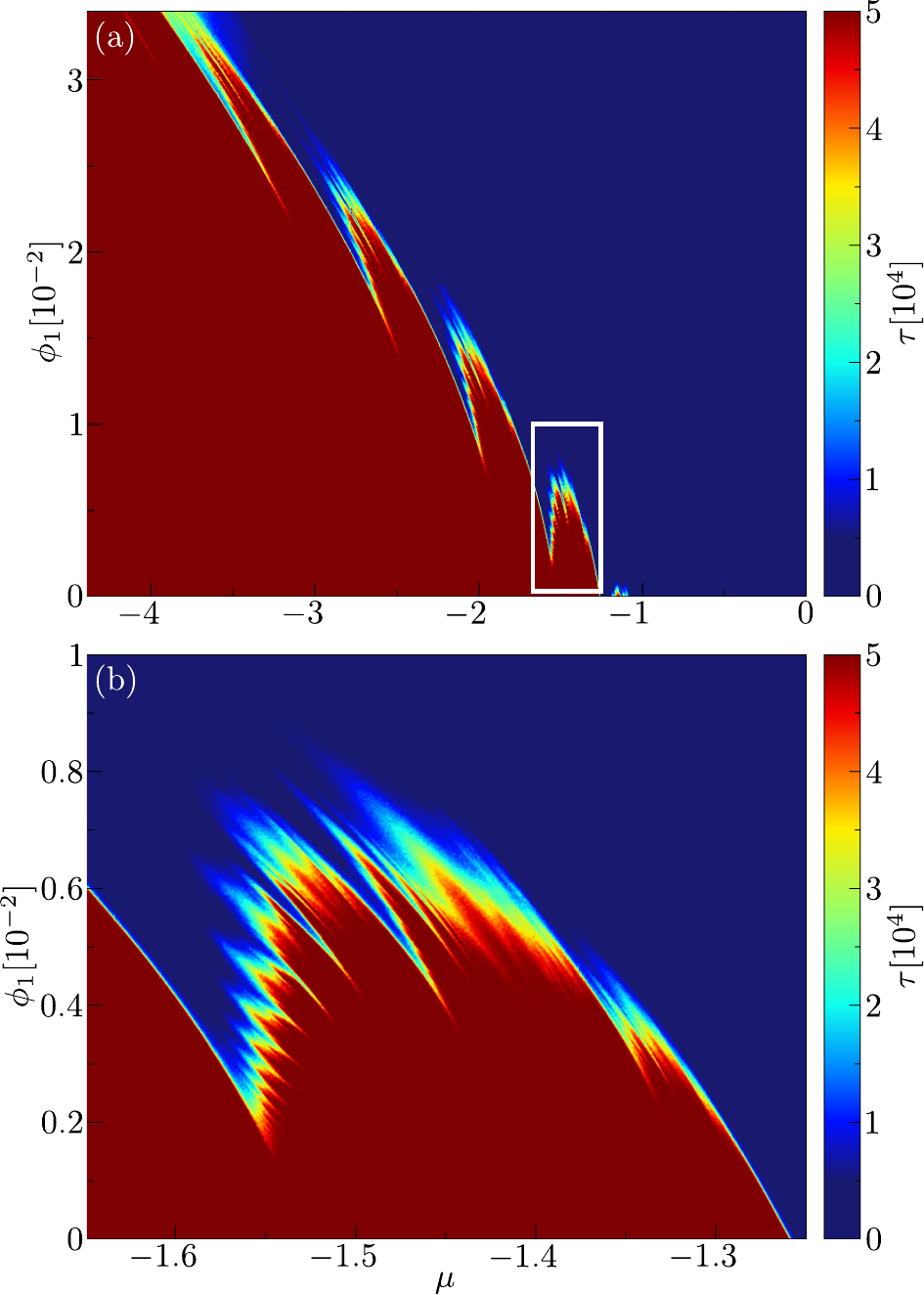}
	\end{center}
	\caption{\label{fig:well_like_Er_Parameter_space} (a) $700\times700$ \textcolor{black}{parameter} space \textcolor{black}{for the well-like radial electric field scenario} with the depth \textcolor{black}{$\mu$} of the electric field and the amplitude \textcolor{black}{$\phi_1$} of the electrostatic potential perturbation. \mbox{(b) Same} resolution magnification inside the white rectangle in panel (a). The colour bar indicates the mean escape time for an ensemble of initial conditions, calculated as shown in equation \eref{eq:Parameter_space_criteria}. \textcolor{black}{We fixed $\eta=\sigma_0$.}}
\end{figure}

\begin{figure}[htb]
	\begin{center}
		\includegraphics[width = 0.48\textwidth]{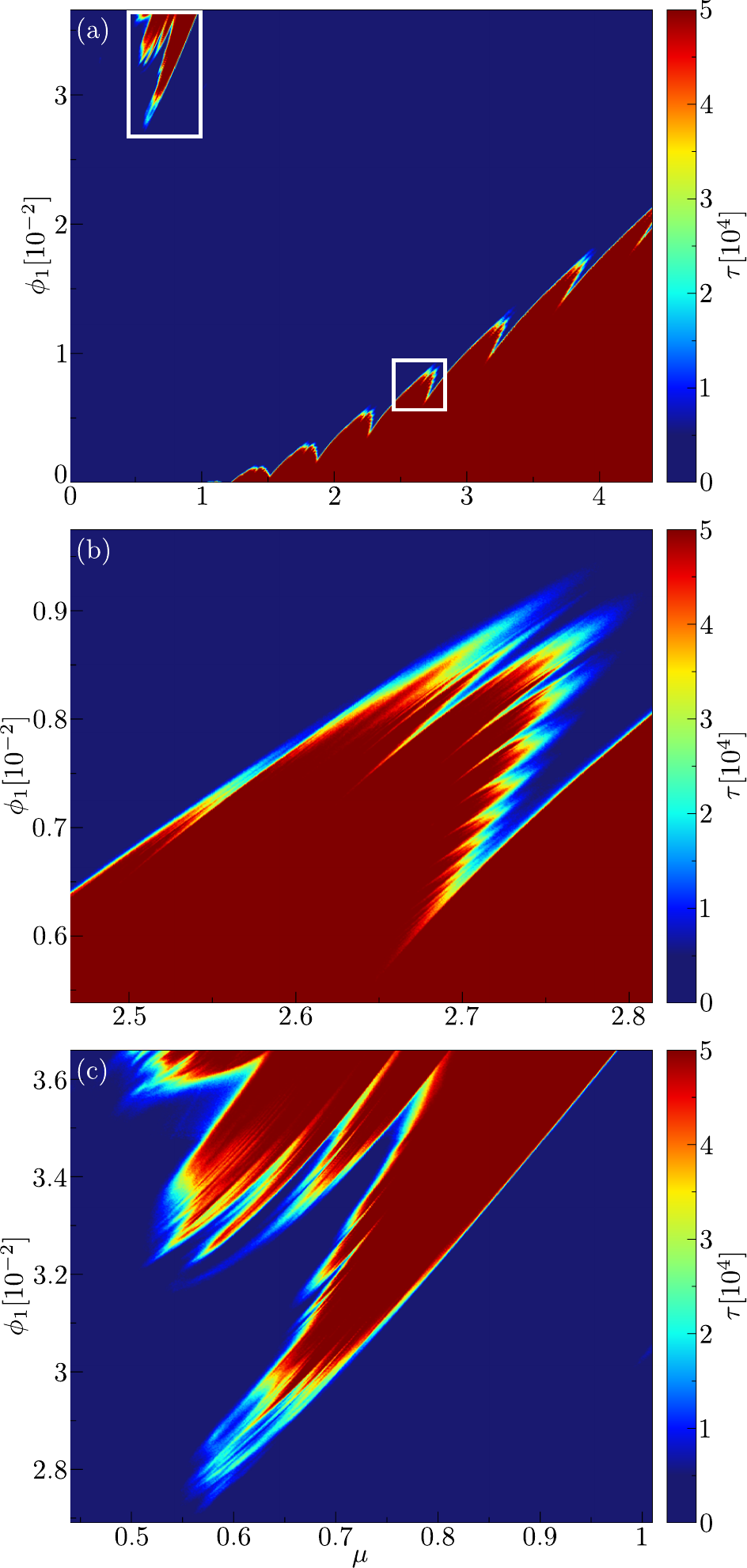}
	\end{center}
 \caption{\label{fig:hill_like_Er_Parameter_space} \textcolor{black}{(a) $700\times700$ \textcolor{black}{parameter} space for the hill-like radial electric field scenario with the height \textcolor{black}{$\mu$} of the electric field and the amplitude \textcolor{black}{$\phi_1$} of the electrostatic potential perturbation. Panels (b) and (c) are magnifications of the same resolution inside the white lower and upper rectangles in panel (a), respectively. The colour bar indicates the mean escape time for an ensemble of initial conditions, calculated as shown in equation \eref{eq:Parameter_space_criteria}. We fixed $\eta=\sigma_0$.}}
\end{figure}

\begin{figure}[htb]
	\begin{center}
		\includegraphics[width = 0.48\textwidth]{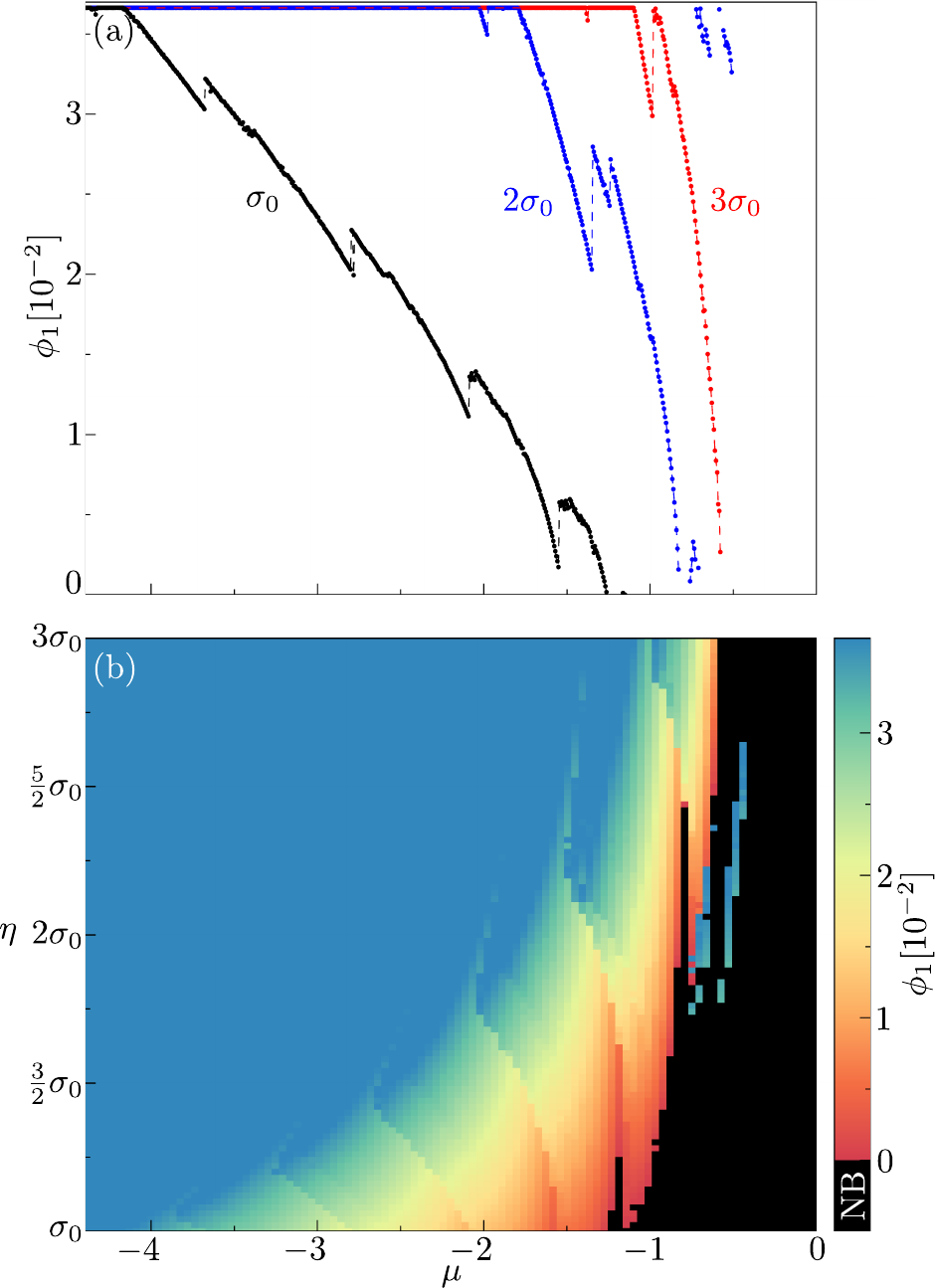}
	\end{center}
	\caption{\label{fig:width_well_like} \textcolor{black}{Well-like $E_\mathrm{H}(I)$ scenario: (a) approximated boundary of the $\mu\times\phi_1$ parameter space for $\eta = \sigma_0$ (black), $2\sigma_0$ (blue) and $3\sigma_0$ (red); (b) maximum value $\phi_1$, up to the precision $\delta\phi_1=5.0\times 10^{-5}$, at which the shearless edge transport barrier exists for the $100\times100$ radial electric field configurations. In black colour, we indicate the cases in which no STB appear.}}
\end{figure}

\textcolor{black}{In the previous section, we found that an STB can appear and confine most of the orbits inside the plasma when the equilibrium radial electric field profile exhibits a pronounced reversed-shear behaviour at the plasma edge. This STB was observed regardless of whether the sheared profile has a positive-to-negative or negative-to-positive shear, corresponding to hill-like or well-like $\textcolor{black}{E_{\mathrm{H}}}(I)$ profiles, respectively.}\par

\textcolor{black}{Now, in this section, we study the robustness of such STBs in terms of the electrostatic potential perturbation amplitude $\phi_1$, which is related to the non-resonant mode $n=1$ in the former case. We explore the parameter spaces of $\mu$, $\eta$ and $\phi_1$ to determine the existence of barriers and evaluate \textcolor{black}{whether} they persist at high values of $\phi_1$. By doing so, we aim to gain insights into the effects of the $\textcolor{black}{E_{\mathrm{H}}}(I)$ profile on the STB robustness and also the chaotic transport at the plasma edge.} \par

\textcolor{black}{So, in first place,} let us consider an ensemble of $\mathcal{P}_{\mathrm{IC}}$ randomly chosen initial conditions in the chaotic region below \textcolor{black}{$I_{\mathrm{edge}}$}\textcolor{black}{, which will be iterated} a maximum of $N_{\mathrm{max}}$ crossings in the Poincar\'e section\textcolor{black}{, and also} a \textcolor{black}{reference} threshold, $I_{\mathrm{up}}$\textcolor{black}{, such that $I_{\mathrm{up}}>I_{\mathrm{edge}}$}. \textcolor{black}{By recording} the time, $\tau_j$, each orbit spends to \textcolor{black}{reach $I_{\mathrm{up}}$,} we can estimate a mean escape time, $\tau$, \textcolor{black}{for a given radial electric field configuration, such that} 

\begin{equation}\label{eq:Parameter_space_criteria}
	\qquad
	\tau = \frac{1}{\mathcal{P}_{\mathrm{IC}}}\sum_{j=1}^{\mathcal{P}_{\mathrm{IC}}}\tau_j,
\end{equation}

\noindent where, for orbits which do not \textcolor{black}{escape, we set $\tau_j$ as being equal to} $N_{\mathrm{max}}$.\par

\textcolor{black}{Then,} when \mbox{$\tau=N_{\mathrm{max}}$}, we will say that, \textcolor{black}{up to the integration time $2\pi N_{\mathrm{max}}/\omega_0$,}  no particle escapes beyond the plasma edge\textcolor{black}{, which, for most of the cases, is related to the onset of a} shearless edge transport barrier. On the other hand, when $\tau<N_{\mathrm{max}}$, there \textcolor{black}{will be no} shearless curve. \textcolor{black}{In some cases, associated with long mean escape times, huge resistances to chaotic transport might appear, such as stickiness regions \cite{szezech2009transport, borgogno2008stable}, which we will refer to as effective barriers from this point henceforth.}\par

\begin{figure}[t]
	\begin{center}
		\includegraphics[width = 0.48\textwidth]{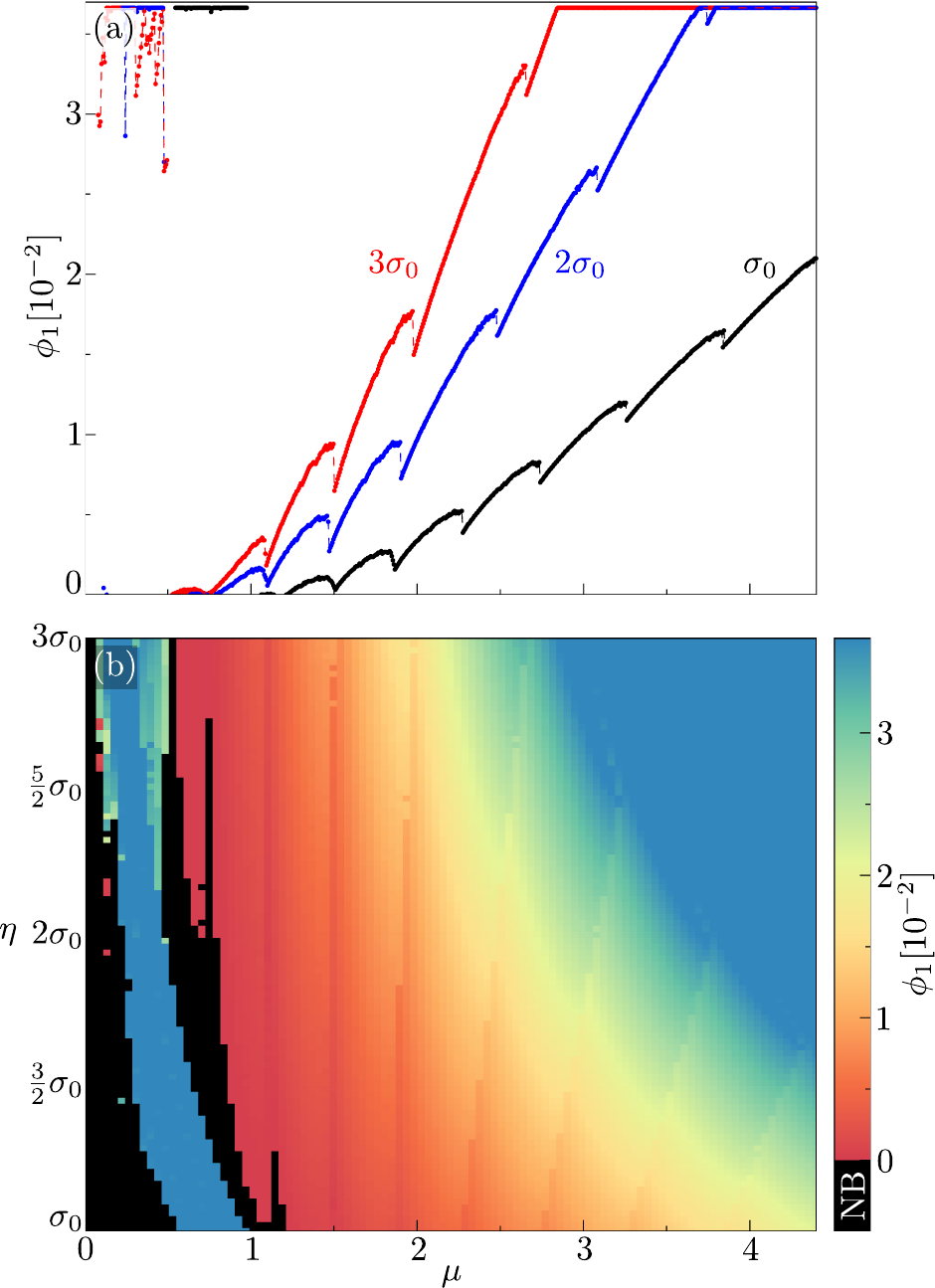}
	\end{center}
	\caption{\label{fig:width_hill_like} \textcolor{black}{Hill-like $E_\mathrm{H}(I)$ scenario: (a) approximated boundary of the $\mu\times\phi_1$ parameter space for $\eta = \sigma_0$ (black), $2\sigma_0$ (blue) and $3\sigma_0$ (red); (b) maximum value $\phi_1$, up to the precision $\delta\phi_1=5.0\times 10^{-5}$, at which the shearless edge transport barrier exists for the $100\times100$ radial electric field configurations. In black colour, we indicate the cases in which no STB appear.}}
\end{figure}

\textcolor{black}{An MPI parallel} code was written to calculate \textcolor{black}{per core} the mean escape time of \textcolor{black}{some edge radial electric field configuration ($\mu,\eta$) perturbed by $\phi_n$, where \mbox{$n=1,2,3$ and $4$}}. We used 352 cores from 11 nodes with the processor Intel Xeon Gold 6142, \textcolor{black}{belonging} to the high-performance computing resources of the Centre de Calcul Intensif d’Aix-Marseille, \textcolor{black}{to compute 23500 electric field scenarios represented in seven figures of parameter spaces (including magnifications), see figures \ref{fig:well_like_Er_Parameter_space}-\ref{fig:width_hill_like}.}\par 

To generate those parameter spaces, we degraded the numerical integrator tolerance to $10^{-9}$ to limit the computational effort, selected $\mathcal{P}_{\mathrm{IC}}=100$ randomly chosen initial conditions in a line at $I=0.7$, and integrated them \textcolor{black}{until they reach the threshold \mbox{$I_{\mathrm{up}}=1.05$}, or until a maximum of $N_{\mathrm{max}}=5\times10^{4}$ crossings in the Poincar\'e section}.\par

\textcolor{black}{With this, in figures \ref{fig:well_like_Er_Parameter_space} and \ref{fig:hill_like_Er_Parameter_space}, we show the $\mu\times\phi_1$ parameter space, for a fixed $\eta=\sigma_0$, to analyze, in the first place, the influence of $\mu$ on the onset, break-up and robustness of the shearless edge transport barriers. From these results,} we notice that, in the \mbox{well-like} $\textcolor{black}{E_{\mathrm{H}}}$ scenario, for small values of $|\mu|$, see \fref{fig:well_like_Er_Parameter_space}(a), no STB or opposition to the chaotic flux outside the plasma appears, \textcolor{black}{all the orbits escape fast} no matter the amplitude of the perturbation. \textcolor{black}{We see an onset of a barrier, associated with $\tau = N_{\mathrm{max}}$, for values of $\mu$ slightly higher than $-1.0$,} however the barrier is broken up easily with a small perturbation. \par 

\textcolor{black}{In the same way, similar results were obtained in the hill-like $\textcolor{black}{E_{\mathrm{H}}}$ scenario, see \fref{fig:hill_like_Er_Parameter_space}(a). Nevertheless, we found that, for small values of $\mu$, the barrier, which has not emerged for small perturbations allowing the escape of orbits, now appears by increasing notably the parameter $\phi_1$. Associated with this barrier, there is a structure on the upper region of the parameter space, which does not cover a great area and has a self-similarity with the whole space.}\par

\textcolor{black}{Now}, as we \textcolor{black}{increase} $|\mu|$, i.e.\ \textcolor{black}{deeper wells or higher hills of the edge radial electric field}, the barrier gains resistance to the perturbation since it breaks up for larger amplitudes of $\phi_1$. It seems that we are accessing improved confinement regimes as the depth \textcolor{black}{(height)} of the electric field well \textcolor{black}{(hill)} increases. \textcolor{black}{So, in that sense, the sign of $\mu$ is not significant in order to reduce particle transport, as identified by \cite{Biglari1990, weynants1992confinement, burrell1999tests}.} \textcolor{black}{\textcolor{black}{Even so}, notice that, in general, for $\mu<0$, larger perturbations are needed in order to break up the STB.} \textcolor{black}{This result seems to be in agreement with \cite{shaing1990bifurcation}, that says the plasma confinement is improved when the radial electric field becomes more negative.}\par

So, in some way, strictly qualitative, we are seeing an L-H transition  through the description of shearless transport barriers which, analogous to the experimental results, exhibit better confinement regimes for larger radial electric fields at the plasma edge.\par 

Furthermore, we notice that, for some windows of the parameter $\mu$, the shearless barrier can appear and disappear recurrently by only varying $\phi_1$, consistently with what has been shown in \cite{marcus2019,osorio2021}. And the same happens if we fix $\phi_1$ and vary $\mu$, as we see clearly from the magnifications shown in figures \ref{fig:well_like_Er_Parameter_space}(b)\textcolor{black}{, \ref{fig:hill_like_Er_Parameter_space}(b) and \ref{fig:hill_like_Er_Parameter_space}(c)}. This parameter space suggests that there is a fractal behaviour, already discovered in other systems with shearless transport barriers, for example, the standard non-twist map \cite{mathias2019fractal}. \par 

Also, it is interesting to notice that effective barriers appear every time the STB breaks up, as we can see by the region with great resistance to the chaotic transport associated with large values of $\tau$. This region covers a larger area of the parameter space when we are in a fractal-like boundary. When we are close to a regular boundary of the parameter space, the area covered by the region is smaller and the huge opposition to the chaotic transport ends easily.\par

\textcolor{black}{Finally, we investigate the effect of the parameter $\eta$, which is related to the width of the profile $\textcolor{black}{E_{\mathrm{H}}}(I)$, on chaotic transport at the plasma edge. For that, we implement an algorithm that, as we did previously, calculates the mean escape time for a given pair $(\mu,\eta)$ varying $\phi_1$ from $\mathrm{max}(\phi_1)$ to $\mathrm{min}(\phi_1)$ with step size $\delta\phi_1$ until obtaining \mbox{$\tau = N_{\mathrm{max}}$}. If we reach \mbox{$\phi_1=\mathrm{min}(\phi_1)$} and $\tau\neq N_\mathrm{max}$, we say \textcolor{black}{that for this} electric field configuration no STB appears. This allows us to estimate the maximum value of $\phi_1$, up to the precision given by $\delta\phi_1$, at which the STB might be found.}\par

\textcolor{black}{Additionally, on varying $\mu$, we can estimate the boundary of the $\mu\times\phi_1$ parameter space for a given $\eta$, as shown in figures \ref{fig:width_well_like}(a) and \ref{fig:width_hill_like}(a), for the associated widths $\sigma_0$, $2\sigma_0$ and $3\sigma_0$, in black, blue and red, respectively. By doing so, we do not need to calculate the entire parameter space for each $\eta$, but rather an approximate boundary. This allows us to investigate the effect of $\eta$ on STBs without excessive computational cost.}\par

\textcolor{black}{Those approximated boundaries are already enough to show that on increasing $\eta$ higher perturbations are needed to break up the transport barriers, for both the well-like and the hill-like scenarios. The width of the electric field profile at the plasma edge is then related to the robustness of the STB. Moreover, we conclude once again that for $\mu<0$ \mbox{more-resistant-to-perturbations} transport barriers can be found than when $\mu>0$, since the boundaries of the $\mu\times\phi_1$ parameter space \textcolor{black}{steepen} more by doubling and tripling the width of the $\textcolor{black}{E_{\mathrm{H}}}$ profile, see figures \ref{fig:width_well_like}(a) and \ref{fig:width_hill_like}(a).}\par

\begin{figure}[htb]
	\begin{center}
		\includegraphics[width = 0.38\textwidth]{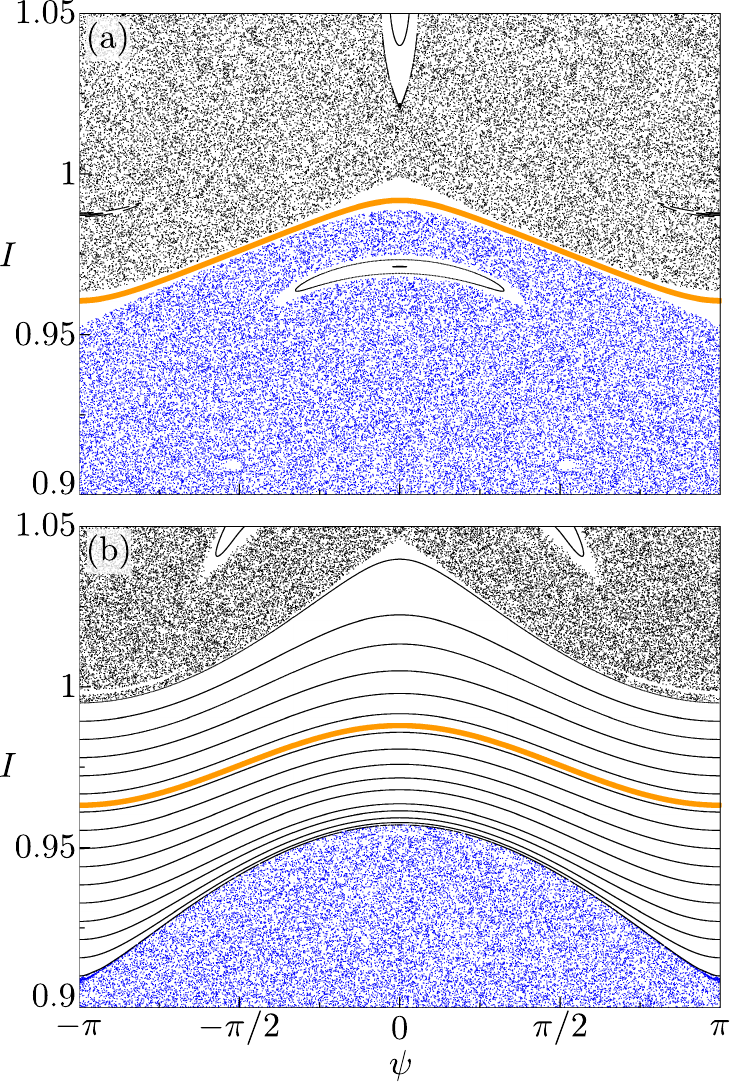}
	\end{center}
	\caption{\label{fig:Phase_space_eta} \textcolor{black}{(a) Magnification at the plasma edge of the Poincar\'e section in \fref{fig:phase_space}(b), where $\eta=\sigma_0$. (b) By increasing $\eta=3\sigma_0$, we suppress chaos in the vicinity of the STB (orange). As a result, higher perturbations are needed in order to allow chaotic transport outside the plasma.}}
\end{figure}

\textcolor{black}{The $\mu\times\eta$ parameter space provides a clearer view of this STB robustness, as seen in figures \ref{fig:width_well_like}(b) and \ref{fig:width_hill_like}(b). These figures illustrate how the barrier is broken up by typically smaller perturbations when $\mu>0$ than when $\mu<0$, and also how it gets robust \textcolor{black}{on} increasing $\eta$ for both types of profiles. Basically, by increasing the width of the electric field at the plasma edge\textcolor{black}{,} we increase the regular region (suppress chaos) in the vicinity of the STB, as shown in \fref{fig:Phase_space_eta}. For that reason, destroying the transport barrier becomes harder as $\eta$ increases in value. In general, this also happens by increasing $\mu$.}\par

\textcolor{black}{Regarding the band shown in \fref{fig:width_hill_like}(b), characterized by high values of $\phi_1$ and small values of $\mu$, it is associated with the structure displayed in \fref{fig:hill_like_Er_Parameter_space}(c). However, further investigation is required to \textcolor{black}{understand} the reasons behind its emergence. Also, notice that a comparable region can be identified in \fref{fig:width_well_like}(b), although it is not as evident.} \textcolor{black}{On the other hand, further progress should be done to verify the effect of additional resonant modes for $n>4$ ($n<0$) on the STBs in the presence of localized strong \mbox{well-like} (\mbox{hill-like}) radial electric fields.}

\FloatBarrier

\section{Conclusions\label{sec:Conclusions}}

\textcolor{black}{Shearless transport barriers (STBs) have been described using an \mbox{$\mathbf{E}\times\mathbf{B}$} drift transport model for a magnetized plasma, considering the electric field as the result of an equilibrium radial part with non-monotonic profile and a perturbation caused by electrostatic fluctuations propagating along the poloidal and toroidal directions. In our model, we also considered radial profiles for the safety factors of the equilibrium magnetic surfaces and the plasma parallel velocity. According to these profiles, a non-monotonic behaviour in the rotation number radial profile can be found, for which the STBs occur at those positions of no shear at all. The numerical simulations presented in this paper were obtained using parameter values taken from the TCABR tokamak, but the results are valid for a wide class of toroidal machines.}\par

\textcolor{black}{This work has provided an analysis of the effect of the electric field radial profile on the emergence of STBs at the edge of a tokamak plasma. We have explored the influence of the intensity and the width of the electric field radial profile on chaotic transport by implementing as diagnostic the mean escape time of an ensemble of particle orbits, which allowed us to characterize the quality of the confinement. In particular, we showed that, due to \mbox{H-mode} radial electric field \mbox{well-like} profiles, STBs can emerge at the plasma edge and may contribute to the decrease of the particle radial flux, thereby improving the plasma confinement. Additionally, we showed that this type of barrier can also be found for hill-like radial electric field profiles.}\par

By shaping the electric field radial profile, we were able to introduce new resonance conditions near the plasma edge, leading to a new dynamics that results in the emergence of a shearless curve which \textcolor{black}{reduces} significantly the particle transport. In particular, this barrier is sensitive to the amplitude of the perturbations, emerging and being destroyed in a recurrent way by the variation of the perturbation amplitudes. With this, we were able to investigate the STB robustness, which is an indirect measurement of the quality of the confinement, in terms of the perturbation strength for several edge radial electric field configurations. \par 

One of our key results is that the STBs become more robust as the depth (height) of the radial electric field well-like (hill-like) profile increases. This is qualitatively in accordance with experimental results about L-H transition. So, the deeper (higher) the electric field well (hill) (i.e.\ the more pronounced the electric field shear), the larger has to be the perturbation strength in order to break up the shearless barrier and have an effective chaotic transport at the plasma edge. We observed a similar behaviour by increasing the width of the profiles. Even so, we found that \mbox{well-like} electric field radial profiles are, in general, related to more-resistant-to-perturbations barriers. \par 

As a numerical diagnostic of the existence of the shearless barrier, we computed the average escape time it takes for a set of guiding-centre orbits to achieve a given threshold above the barrier location. If this escape time reaches its maximum value, we say that, up to the numerical accuracy, no particle escapes and, therefore, the shearless barrier exists. The appearance and disappearance of barriers were found to depend on the control parameters in an intermittent fashion, where, from the two-dimensional parameter spaces surveyed, we observed a frontier \mbox{barrier-non-barrier} transiting between a fractal behaviour and a regular one. Moreover, this technique also allowed us to characterize parameter space regions of effective confinement or effective barrier behaviour, occurring when the particles spend a long time in the plasma before escaping away (there is not an STB). We found that these regions exist before the appearance of the STB and after its disappearance, no matter how small the transition interval is considered.\par    

\textcolor{black}{In conclusion, our findings suggest that both the intensity and the width of the radial electric field profile play a crucial role in promoting a more-robust STB, which may contribute to the decrease of the particle radial flux at the plasma edge. These results highlight the importance of carefully shaping the edge radial electric field profile to achieve optimal confinement in fusion devices.}\par

\ack The authors thank the financial support from the Brazilian Federal Agencies (CNPq), grants 407299/2018-1, 302665/2017-0, \textcolor{black}{403120/2021-7} and \textcolor{black}{301019/2019-3}, the S{\~a}o Paulo Research Foundation (FAPESP, Brazil) under grants 2018/03211-6, 2018/14435-2, 2020/01399-8 and \textcolor{black}{2022/04251-7}, the Coordena\c{c}\~{a}o de Aperfei\c{c}oamento de Pessoal de N{\'i}vel Superior (CAPES) under grants 88881.143103/2017-01 and 88887.675569/2022-00, and the  Comit{\'e} Fran\c{c}ais d'Evaluation de la Coop{\'e}ration Universitaire et Scientifique avec le Br{\'e}sil (COFECUB) under grant No. 40273QA-Ph908/18.\par

The Centre de Calcul Intensif d'Aix-Marseille is acknowledged for granting access to its high-performance computing resources. 

\section*{References}
\bibliography{biblio.bib}

\end{document}